\documentstyle[preprint,aps,epsf,rotate]{revtex}
\begin{document}
%
\preprint{NSF-ITP 97-042}
\title{Critical Behavior of Random Bond Potts Models}
\author{John Cardy$^{1,2}$ and Jesper Lykke Jacobsen$^{1,3}$}
\address{$^1$University of Oxford, Department of Physics -- Theoretical
         Physics, \\ 1 Keble Road, Oxford OX1 3NP, U.K. \\
         $^2$All Souls College, Oxford; $^3$University of Aarhus, Denmark.}
%
%
\maketitle
\begin{abstract}
The effect of quenched impurities on systems which undergo
first-order phase transitions is studied within the framework of the
$q$-state Potts model. For large $q$ a mapping to the random field Ising
model is introduced which provides a simple physical explanation
of the absence of any latent heat in 2D, and suggests that
in higher dimensions such systems should exhibit a tricritical point
with  a correlation length  exponent related to the
exponents of the random field model by
$\nu=\nu_{\rm RF}/(2-\alpha_{\rm RF}-\beta_{\rm RF})$.
In 2D we
analyze the model using finite-size scaling and conformal invariance, and
find a continuous transition with a magnetic exponent $\beta/\nu$ which
varies continuously with $q$, and a weakly varying correlation length
exponent $\nu\approx1$. We find strong evidence for the multiscaling of the
correlation functions as expected for such random systems.
\end{abstract}
\pacs{PACS numbers: 05.70.Jk, 64.60.Ak, 64.60.Fr}
%
%
Although the effect on the critical behavior of adding quenched bond
randomness to 
classical systems whose pure version undergoes a continuous phase
transition is well understood in terms of the Harris criterion
\cite{Harris}, the
analogous situation when the pure transition is first order is less
well studied. Following earlier work of Imry and Wortis \cite{IW}, 
Aizenman and Wehr \cite{AW} and Hui and Berker \cite{HB} argued
that in 2D any amount of randomness
should lead to a vanishing of the latent heat.
The arguments leading to this conclusion are analogous to those used by
Imry and Ma \cite{IM}
for the absence of any spontaneous magnetization in the random
\em field \em Ising model (RFIM) for $d=2$: the bond randomness couples to the
local energy density, which is different for the coexisting phases of
the pure model, in the same manner that the random field couples to the
local magnetization of the RFIM. The vanishing of the latent heat
should be accompanied by a divergent correlation length, and, if so, the
question arises as to which universality class(es) the corresponding
continuous transition belongs. A suitable model in which to study this
is the $q$-state Potts model, whose pure version in 2D undergoes a 
first-order transition for $q>4$, otherwise being continuous. Chen,
Ferrenberg and Landau \cite{CFL} undertook an extensive Monte Carlo
investigation of the case $q=8$. In addition to confirming the
continuous nature of the transition, these authors extracted numerical
values of the critical exponents which appear to be consistent 
with those of the \em pure \em 2D Ising model.
Similar values have also been claimed for the case $q=4$, when the pure
transition is continuous \cite{DW}. This disagrees with the predictions of
Ludwig and Cardy \cite{Ludwig87}, Ludwig \cite{Ludwig90} and Dotsenko 
{\em et al.}~\cite{Dotsenko},
who find a new random fixed point for $q>2$, based on an expansion 
in powers of $q-2$.

In this Letter we describe the results of an investigation of this model
along two different lines. First, we show that at large $q$ the model
(or rather a model for the interface between the disordered and ordered
phases) may be mapped onto a corresponding problem for the RFIM. This
mapping gives a direct and intuitive picture for the arguments of
Refs.~\cite{AW,HB}. Moreover, since the RFIM interface
is well understood from the renormalization group (RG) perspective
\cite{RFIM}, one may
immediately transcribe these results to give quantitative
predictions for the random bond problem, both in two and in higher
dimensions. The second line uses the powerful
techniques of conformal invariance, combined with finite-size scaling,
which should apply to such a system with a divergent correlation length.
By generalizing the transfer matrix formalism of Bl\"ote and Nightingale
\cite{Blote}
we are able to generate results for arbitrary $q$ and reasonably large
strip widths. At the same time we have to take account of the fact that
in random systems the transfer matrices do not commute, and one must
discuss the spectrum of Lyapunov exponents rather than the eigenvalues.
The relation of these to critical exponents is indirect, but we have
developed a method of making this connection based on a cumulant
expansion, in which we see explicitly the multiscaling properties of the
correlation functions discussed theoretically by Ludwig \cite{Ludwig90}.
The above method works poorly for
the thermal exponent. Instead, by finite-size scaling arguments
\cite{Nightingale} we are able to determine $\nu$ directly.

As will become clear, many of our results generalize, but let us for
definiteness consider a Potts model on the square lattice with
degrees of freedom $s_i$ taking $q$ values, and a reduced hamiltonian
$-\sum_{ij}K_{ij}\delta_{s_is_j}$, where the sum is over nearest
neighbor pairs. The ferromagnetic couplings $K_{ij}$ are
quenched random variables, taking the values $K_1$ and $K_2$, each with
probability $\frac12$. When $(e^{K_1}-1)(e^{K_2}-1)=q$ this model
is, on average, self-dual, and, if the transition is unique, is therefore
at its critical point \cite{Kinzel}. It is useful to parametrize
$e^{K_{ij}}-1=u_{ij}=q^{\frac12+w_{ij}}$, where
$w_{ij}=\pm w$, and $w>0$ measures the strength of
the randomness. The partition function of this model may be mapped
onto that of the random cluster model \cite{Kasteleyn},
in which each bond of the lattice is either
occupied, when it is counted with weight $u_{ij}$, or
empty, in which case it is counted with weight 1. The partition sum is over all
such configurations, in which each connected
cluster of sites is weighted by a factor $q$.
Let us first consider the pure model, with $w=0$. In the limit
$q\to\infty$, the sum over configurations is dominated by only two: the
\em empty \em lattice, in which no bonds are occupied, which contributes
a factor $q^N$, where $N$ is the total number of sites, and the
\em full \em lattice, with a weight $(\sqrt q)^{2N}$, since the number
of bonds is $O(2N)$. All other
configurations are down by powers of $q^{\frac12}$. At the self-dual
point, there are therefore two co-existing states with identical bulk free 
energy and different internal energy densities, indicating, as expected, that
the transition has a non-vanishing reduced latent heat per bond
$\sim\frac12\ln q$. For the pure model, this analysis may be extended to take
into account higher order corrections in $q^{-\frac12}$, with no
essential change in the physical picture. Now consider an \em interface
\em between these two phases. For large $q$, 
the lowest energy interface is parallel to
a lattice direction, say the $x$-axis, and is such that all the bonds
with $y\leq$ some integer are occupied, and those above this are empty (or
vice versa.) There will also be entropic fluctuations 
$y=h(x)$ of this interface, described by the usual solid-on-solid
interfacial hamiltonian, proportional to the length of the interface. 
The interfacial tension for large $q$
is $\sigma\sim\frac14\ln q$, independent of the local shape of the interface.
This is to be compared with $\sigma\sim2J$
between the \em ordered \em phases of a low temperature Ising model
with reduced exchange coupling $J$. 

Now consider the effect of adding bond randomness to the random
cluster model. Each configuration of the interface will be weighted by
an energy $\sum_x\sum_{y<h(x)}{w(x,y)}\ln q$, where $(x,y)$ labels
bond positions. This may be rewritten, up to a term independent of
$h(x)$, as 
$\frac12\sum_x\big(\sum_{y<h(x)}-\sum_{y>h(x)}\big)w(x,y)\ln q$.
Compared with the energy of an RFIM interface
with spins $s(x,y)=\pm1$ coupled to a reduced random field
$h(x,y)=\pm h_{\rm RF}$,
the interfacial models are identical with the
correspondence $J\leftrightarrow\frac18\ln q$, and
$h_{RF}\leftrightarrow\frac12w\ln q$. In addition, the imposition of a
uniform reduced magnetic field $h$ on the RFIM, which
distinguishes between the two co-existing phases, is seen to be
equivalent to a deviation $t\equiv (T-T_c)/T_c$
in the \em temperature \em variable away from
the critical self-dual point. Since this couples to the energy density
we find the correspondence $h\leftrightarrow\frac14t\ln q$.

Of course, this is strictly valid only as $q\to\infty$.
At finite $q$ the $q$-dependence of
cluster configurations with more complicated topologies 
is not simply accounted for by the interfacial tension. 
For the same reason, the mapping is not between \em bulk \em
configurations of the two models.
However, it will be argued that certain universal properties
are controlled by an
RG fixed point at infinite $q$, and for these the mapping
should be asymptotically exact. Although this
has been described in terms of a 2D self-dual model, it
should be clear that it is more general: lack of
self-duality corresponds to a skewness in the distribution of the
random fields $h(x,y)$, which may be compensated by adding a suitable
uniform field (corresponding to a shift in the $T_c$ of the Potts
model), and, similarly, higher dimensions may be taken into account by
appropriately replacing $\sqrt q$ by $q^{1/d}$. 

\begin{figure}[t]
  \centerline{\epsfxsize=4.5in \epsfbox{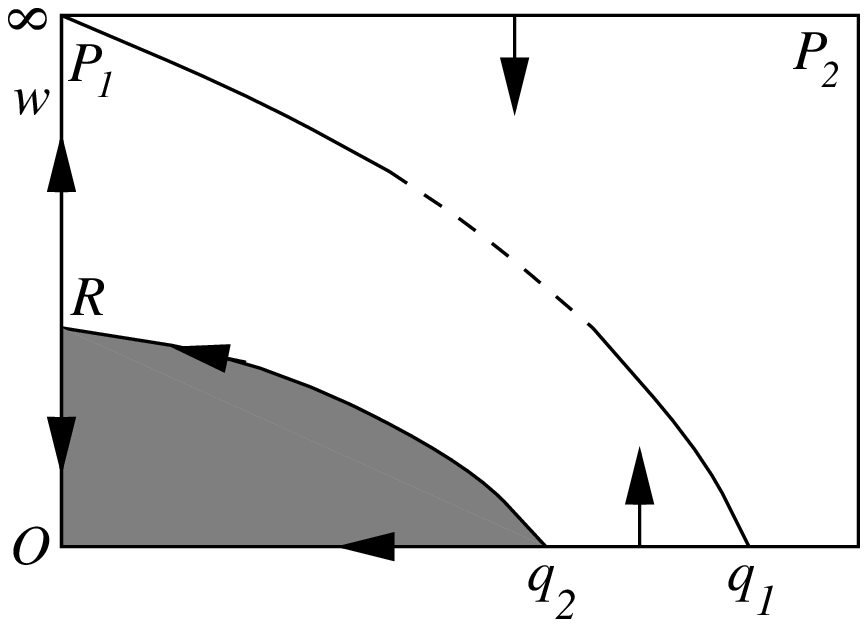}}
  \caption{Schematic phase diagram in the critical surface for $d>2$. $q$
increases to the left and $w$ is the disorder strength, with $P_1P_2$ being
the percolation limit. RG flows are indicated.
The latent heat is non-vanishing within the shaded region, and
elsewhere the transition is continuous, controlled by the line of fixed
points $P_1q_1$.} 
  \label{fig-pd}
\end{figure}

The RG properties of the interface in the RFIM near 
$d=2$ have been well studied \cite{RFIM}. When translated
into the variables of the random bond Potts model, the flow equations
have the form
\begin{eqnarray}
dw/dl&=&-(d/2-1)w+Aw^3+\cdots\label{rg1}\\
d(\ln q)^{-1}/dl&=&-(\ln q)^{-1}\big((d-1)-Aw^2+\cdots\big)\label{rg2}\\
dt/dl&=&t(1+Aw^2+\cdots),\label{rg3}
\end{eqnarray}
where $A>0$ is a non-universal constant. Corrections to
these equations are supposed to be higher order in $w$ and in
$q^{-1/2}$. The reader who is concerned that the number of Potts states
$q$ is allowed to flow in these equations is reminded that $q$ is merely
a parameter of the random cluster model. 
The RG flows for $d>2$ and the consequent phase diagram are shown in
Fig.~\ref{fig-pd}.
In the pure models, for $q>$ some $q_2(d)$ (low $T$ in the RFIM),
there is phase coexistence with a non-vanishing latent heat
(spontaneous magnetization), controlled by a fixed point at infinite $q$
($T=0$). For $d>2$ this persists into the shaded region, bounded by a line
of tricritical points where the
latent heat, vanishes. The universal behavior 
along this line is controlled by the fixed point $R$ at 
$w=O\big((d-2)^{1/2}\big)$ and infinite $q$. 
Using the correspondence
$t\leftrightarrow \frac{1}{2} h\cdot T_{\rm RF}$, we
conclude that the thermal eigenvalue $\nu^{-1}$ of the random bond
problem is related to the RG eigenvalues and exponents
of the RFIM by the scaling
relations $\nu^{-1}=y_h-\theta=(2-\alpha_{\rm RF}-\beta_{\rm RF})/\nu_{\rm RF}$,
where $-\theta$ is
the eigenvalue which controls the irrelevance of $T_{\rm RF}$
and the consequent violation of hyperscaling \cite{RFIM}. 
In the same manner, it may be shown that the latent heat vanishes 
as $(w_c-w)^{\beta_{\rm RF}}$ as the line $Rq_2$ is approached from below.
Of course, these relations have been established only close to $d=2$, but, if
the topology of the RG flows does not change, they should hold also in
three and higher dimensions.

Above the line $Rq_2$, the flows 
go to large $w$ beyond the validity of (\ref{rg1}-\ref{rg3}). In addition,
the renormalized
interfacial tension flows to zero and the mapping between the models 
breaks down as domains of different topologies proliferate.
However, for infinite $q$
the mapping remains exact and the flows go to infinite $w$.
This cannot happen for finite $q$ since this is the percolation limit
$K_1/K_2=0$, at which $w^{-1}$ is \em relevant \em\cite{percolation}. There
must therefore exist another 
line of stable fixed points emerging from $P_1$, which
control the universal continuous transition for large, but finite, values
of $w$ and $q$. It is tempting to conjecture, as indicated by the dashed line
in Fig.~\ref{fig-pd}, that this connnects on to that found by
expansion in powers of $q-q_1$ \cite{Ludwig87}, where $q_1$ is the point where
the exponent $\alpha$ of the pure model changes sign \cite{Harris}.
Our analysis indicates that, at least for $d=2$, this is the case. 
In 2D (when $q_1=2$, $q_2=4$), the shaded region collapses,
and for any non-zero $w$ the renormalized interfacial tension, and thus
the latent heat, vanish. The flows should be towards the line $P_1q_1$,
with a cross-over length which, from Eq.~(\ref{rg1}), has the form 
$\xi_X\sim e^{1/2Aw^2}$ and therefore may become very large for weak 
randomness.

We now turn to our numerical results for $d=2$. These will be described in 
detail elsewhere \cite{JLJ}. 
As shown by Bl\"ote and Nightingale \cite{Blote} the transfer
matrix for the pure $q$-state Potts model in a strip of width $L$ may be
constructed in a basis in which $q$ enters only as a continuous parameter.
We have generalized this to the case when the bond strengths $u_{ij}$ are 
quenched random variables, and the transfer matrices ${\cal T}_i$ therefore
depend on the row labels $i$. The size of the transfer matrices grows like
$4^L$, independent of $q$, making this a practicable method for larger $q$.

\begin{figure}[t]
  \centerline{\epsfxsize=4.5in \epsfbox{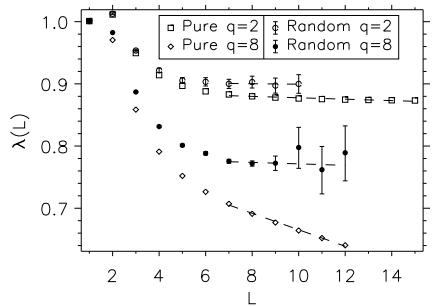}}
 \caption{Plots of $\lambda (L)$, normalized to $\lambda(1)=1$, showing that
           bond randomness renders the phase transition second order (see 
           text).}
  \label{fig-12}
\end{figure}

Starting with some suitable initial vector
${\bf v}_0$, the leading Lyapunov exponent is given by
\cite{Benettin} 
$\Lambda_L^0=\lim_{m\to\infty}\frac1m\ln||
\left( \prod_{i=1}^m{\cal T}_i \right) {\bf v}_0||$.
Higher exponents are found by iterating a set of vectors
$({\bf v}_i)_{j=0}^k$, where a given ${\bf v}_j$ is orthogonalized
to the set $({\bf v}_i)_{i=0}^{j-1}$ after each multiplication by the
transfer matrix.
The average free energy per site is then $f_L=-\frac1L\Lambda_L^0$. For a 
system exhibiting a first-order transition with a bulk correlation length
$\xi$ we expect \cite{Blote} $f_L\sim f_\infty+O(L^{-d}e^{-L/\xi})$, so that
$\lambda(L)\equiv\ln[f_L-f_\infty]+d\ln L\sim{\rm const}-L/\xi$. 
In Fig.~\ref{fig-12} we show this for various values of $q$ and the randomness
strength $R\equiv K_2/K_1$. We see that the randomness changes the transition
for $q=8$ into one with an apparently diverging correlation length. In such a 
case the amplitude of the finite-size correction has the form \cite{BCN}
$f_L\sim f_\infty-\pi c'/6L^2$, where $c'$ is the \em effective \em
central charge (since $f_L$ is the quenched free energy, in a replica
formalism this is the derivative of the central charge $c(n)$ wrt the number
of replicas at $n=0$). The value of $c'$ was determined by making parabolic
least-squares of $f_L$ versus $1/L^2$ \cite{Queiroz95}. 
We found the optimum trade-off between
statistical errors  and a reasonable computation time by taking strips of 
length $m=10^5$ and averaging $f_L$ over $100$ independent realizations of
the randomness for $1\leq L\leq8$, and $3$ realizations for $9\leq L\leq12$.
Data collection was made for each 200 multiplications by ${\cal T}_i$, and the 
first 2,000 iterations of each run were discarded in order to eliminate
transients. The parabolic fits were made by including the data points for
$L_0\leq L\leq12$, where $L_0$ must be chosen large enough to justify the
finite-size scaling form, and small enough to minimize error bars. From the
special cases of the Ising model and percolation it appeared that $L_0=3$
is optimal.

For the random bond Ising model ($q=2$) with $R=2$ we found $c'=0.495\pm0.006$,
in agreement with de Queiroz' result \cite{Queiroz95}
$c'=0.498\pm0.003$ using the spin basis,
and with the expected value $c'=\frac12$. For $q=3$ and $R=2$, our
result $c'=0.799\pm0.006$ is unable to distinguish between the pure value
of $\frac45$ and that of $c'\approx0.8025$ obtained in Ref.~\cite{Ludwig87}
by an expansion in $q-2$. For $q=4$ the results $c'=1.003\pm0.006$ for $R=2$
and $c'=1.010\pm0.022$ for $R=10$ are consistent with each other and the
pure value
$c=1$, but for larger $q$ the $R=2$ results appear to saturate while those
for $R=10$ show a gradual increase: $c'=1.517\pm0.025$ for $q=8$ and
$c'=3.003\pm0.031$ for $q=64$. Similar values have recently been reported
by Picco \cite{Picco}. However, it should be pointed out
that these are also very close to those expected at the percolation point
$R\to\infty$. For then the replicated model is the Potts model with $q^n$
states, so that $c'=(\partial/\partial n)c_{\rm Potts}(q^n)|_{n=0}=
(5 \sqrt 3/4\pi)\ln q\approx 0.689\ln q$; this is confirmed by our
transfer matrix calculations \cite{JLJ}. These are also remarkably
close to the pure values for $2\leq q\leq4$. We conclude that
measurement of the effective central charge does not distinguish well between
pure, percolative and non-trivial random behavior.

\begin{figure}[t]
  \centerline{\epsfxsize=4.5in \epsfbox{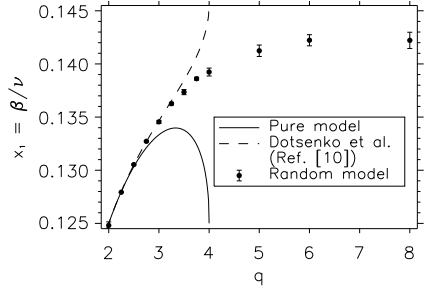}}
  \caption{Magnetic exponent $x_1=\beta/\nu$ as a function of $q$.}
  \label{fig-x1}
\end{figure}

We therefore turned to the magnetic exponent $x_1=\beta/\nu$. Although this
may be determined in principle by adding a ghost site \cite{Blote}, it proved
more efficient to exploit duality and relate the spin-spin correlation
function $G(m)$ on the strip to the free energy in the presence of a seam
of frustrated bonds. Details of this relation and how to implement it in
the transfer matrix will be given elsewhere \cite{JLJ}. In pure systems,
according to conformal symmetry \cite{C2}
$G(m)$ decays along the strip as $\exp(-2\pi x_1m/L)$, so that
the difference $\Delta f_L$ 
in the free energy per site with and without the seam
behaves as $2\pi x_1/L^2$. However, in the random system this difference
typically has fluctuations which are $O(m^{-1/2})$. This has the consequence
that, while $\Delta f_L\sim \frac1{mL}\ln G$ is self-averaging, 
$G$ is not \cite{Derrida}. 
In fact, as shown by Ludwig \cite{Ludwig90}, the moments $\overline{G(m)^N}$ 
(where the overline denotes the quenched average) exhibit
\em multiscaling\em, that is, they scale with dimensions $x_N$ which are not,
in general, linear in $N$. 
Since conformal symmetry assumes translational invariance, it refers only to
such averaged quantities. In practice, we
can avoid the lack of self-averaging by performing a
cumulant expansion 
\begin{equation}
\ln\overline{G^N}=N\overline{\ln G}+\textstyle{1\over2}
\displaystyle N^2\overline{(\ln G-
\overline{\ln G})^2}+\cdots,
\end{equation}
where the higher cumulants may be directly extracted from the statistical
fluctuations in $\Delta f_L$. For values of $q$ and $R$ which are not too
large, this expansion appears to converge well, keeping the first 3 or 4 
cumulants. The fact that the higher cumulants are non-zero implies
multiscaling. Our values for $x_1$ are shown in Fig.~\ref{fig-x1}. For $2<q\leq
3$ and $R=2$ they are in perfect agreement with the predictions of 
the $(q-2)$-expansion of Refs.~\cite{Ludwig90,Dotsenko}. 
For larger $q$ the 
results appear to increase smoothly with $q$, with $x_1(q=8)=0.142\pm0.001$.
Thereafter the cumulant expansion begins to break down. 

\begin{figure}
  \centerline{\epsfxsize=4.5in \epsfbox{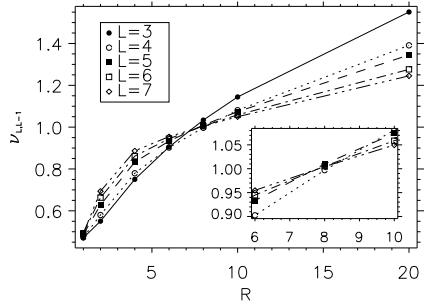}}
  \caption{Values of $\nu$ extracted from phenomenological RG for different
           strip widths and $q=8$. There is a fixed point at $R\approx8$ 
           with $\nu\approx1$ (see inset). Error bars are less than the
           symbol size.}
  \label{fig-nu}
\end{figure}

Although the thermal exponent $\nu$ should be related in a similar manner
to the first gap $\Lambda_L^1 -\Lambda_L^0$ in the Lyapunov spectrum, for
reasons we do not understand this yields results which, if
taken literally, appear to violate the bound $\nu\geq1$ \cite{Chayes}. 
Instead we have measured $\nu$ directly by finite-size scaling of the
magnetic correlation length away from $T_c$, using phenomenological
RG methods \cite{Nightingale}.
For $q=8$, as shown in Fig.~\ref{fig-nu}, we find
clear evidence for fixed points at $R=0$, with
$\nu=\frac12$ ($x_T=0$), and a random fixed point with
$\nu_{\rm R} = 1.01 \pm 0.02$. At $q=3$ our results are consistent with
the perturbative value $\nu_{\rm R} \approx 1.02$, and for $q=64$ we find
$\nu_{\rm R} = 1.02 \pm 0.03$. 

In summary, we have computed the exponents of the 2D random bond Potts model
and shown that while the thermal exponent $\nu_{\rm R}$ is consistent,
within error bars, with both
the pure Ising value, and with the results of the 
$(q-2)$-expansion, the magnetic exponent varies continuously with $q$ 
in a manner which agrees with this expansion in the region it is expected
to be valid, and
with a value at $q=8$ which is quite different from the Ising value of 
$\frac18$. This is in sharp disagreement with the Monte Carlo results of
Ref.~\cite{CFL}. One possible reason is that these 
authors use a non-standard definition of the order parameter which, while
it scales in the same way as the usual one in the pure case, may not when
multiscaling is present. Another is that our very long strips are able to
accommodate large regions in which all $q$ values of the order parameter are 
realized, and 
this may not typically be the case in the square geometries of Ref.~\cite{CFL}.
We have also given a mapping of the random bond problem to the RFIM which 
suggests that for $d>2$ such systems should exhibit a tricritical point whose
thermal exponent is related to those of the RFIM. This picture is
quite generic, and, since there are many real 3D systems which undergo
first-order transitions, it would be interesting to re-examine the effect
of random impurities in such cases. 

The authors acknowledge useful discussions with J.~Chalker, A.~Ludwig and
S.~de Queiroz. This research was supported in part by the Engineering and 
Physical Sciences Research Council under Grant GR/J78327, and by the 
National Science Foundation under Grant PHY94-07194.


\end{document}